\documentstyle[12pt,psfig]{article}
\def\beq{\begin{equation}}
\def\eeq{\end{equation}}
\def\bea{\begin{eqnarray}}
\def\eea{\end{eqnarray}}
\def\bq{\begin{quote}}
\def\eq{\end{quote}}

\def\gappeq{\mathrel{\rlap
{\raise.5ex\hbox{$>$}}
{\lower.5ex\hbox{$\sim$}}}}

\def\lappeq{\mathrel{\rlap{\raise.5ex\hbox{$<$}}
{\lower.5ex\hbox{$\sim$}}}}
\def\simlt{\stackrel{<}{{}_\sim}}
\def\simgt{\stackrel{>}{{}_\sim}}

\begin{document}
\pagestyle{empty}
\begin{flushright}
{CERN-TH/98-328\\
 IFT-98/21\\
TPI-MINN-98/21\\
 UMN-TH-1726/98\\
hep-ph/9811284}
\end{flushright}
\vspace*{5mm}
\begin{center}
{\bf
COSMOLOGICAL FINE TUNING, SUPERSYMMETRY  AND THE GAUGE HIERARCHY PROBLEM}
\\
\vspace*{1cm} 
Piotr H. Chankowski$^{a)}$, John Ellis$^{b)}$, 
Keith A. Olive$^{c)}$ and Stefan Pokorski$^{a)}$
\\
\vspace*{1.7cm}  
{\bf ABSTRACT} \\
\end{center}
\vspace*{5mm}
\noindent
We study the extent to which the cosmological fine-tuning problem - 
why the relic density of neutralino cold dark matter 
particles $\chi$ is similar to that of baryons - is related 
to the fine-tuning aspect of the gauge hierarchy problem -
how one arranges that $M_W \ll M_P$ without unnatural choices of
MSSM parameters. Working in the minimal supergravity framework
with universal soft supersymmetry-breaking parameters as inputs, we find
that the hierarchical fine
tuning is minimized for $\Omega_{\chi} h^2 \sim 0.1$. Conversely, imposing
$\Omega_{\chi} h^2 < 1$ does not require small hierarchical fine tuning,
but the exceptions to this rule are rather special, with parameters
chosen such that 
$m_{\chi}\sim M_Z/2$ or $M_h/2$, or else $m_{\chi} \simgt m_t$. In 
the first two cases, 
neutralino annihilation receives a large contribution from a 
direct-channel pole, whereas in the third case
it is enhanced by the large top Yukawa coupling.
\vspace*{1.0cm} 
\noindent

\rule[.1in]{16.5cm}{.002in}

\noindent
$^{a)}$ Institute of Theoretical Physics, Warsaw University, Poland \\
$^{b)}$ Theory Division, CERN, Geneva, Switzerland \\
$^{c)}$ Theoretical Physics Institute,
School of Physics and Astronomy, University of Minnesota,
Minneapolis, MN 55455, USA \\
\vspace*{0.2cm}

\begin{flushleft} CERN-TH/98-328\\
IFT-98/21 \\
TPI-MINN-98/21\\
UMN-TH-1726/98\\
November 1998
\end{flushleft}
\vfill\eject
%\pagestyle{empty}
%\clearpage\mbox{}\clearpage
\newpage

\setcounter{page}{1}
\pagestyle{plain}

One of the important philosophical issues to be addressed by any dark
matter candidate is the extent to which a relic density of interest
to astrophysicists and cosmologists, $0.01 < \Omega h^2 < 1$, is
natural. Indeed, this question is often posed to proponents of
favoured candidates such as a neutrino, the axion and the lightest
supersymmetric particle, assumed to be the lightest neutralino $\chi$
\cite{mark}. In the case of a neutrino, the see-saw mechanism explains in
a natural way why $m_{\nu} \ll m_q$ or $m_{\ell}$, but does not lead
inexorably to $\Omega_{\nu}$ in the interesting range. In the case of the
axion, experimental and astrophysical constraints restrict its relic
density to the interesting range, but this is not yet predicted by any
deeper theoretical argument, although such an argument may yet be found
\cite{raffelt}. In the case of supersymmetry on the other hand, it is
well known and will be emphasized below that there are indeed good
physical reasons for expecting cosmologically significant relic
densities \cite{jkg}.

The essential motivation for supersymmetric particles to appear below the
TeV scale is provided by the gauge hierarchy problem. Supersymmetry by
itself does not explain why $M_W \ll M_P$, but it does enable such a
hierarchy to be stabilized against the effects of radiative corrections,
averting the need for fine tuning and rendering the gauge hierarchy
technically natural~\cite{naturalness}. The appearance of supersymmetric
particles at the
TeV scale is also supported \cite{SGUT,CHPLPO} by the experimental value of 
$\sin^2\theta_W$, in accord with supersymmetric grand unified theories, 
and by the indications that the Higgs boson may weigh around
100~GeV~\cite{hundred},
in agreement with supersymmetric model calculations if sparticles
weigh $< 1$~TeV~\cite{susyhiggsmass}. 

The TeV scale also arises as a
possible
characteristic mass scale for a cold dark matter candidate. Particles
that annihilate via conventional point-like interactions generically
have $\Omega \sim 1$ if their masses $m \sim \sqrt{M_P \times T_{CMBR}}$,
where the cosmic microwave background temperature $T_{CMBR} =
2.73~K$~\cite{Dim}. It
so happens that $\sqrt{M_P \times T_{CMBR}} \sim 1$~TeV, making it
plausible that any relic with mass around the electroweak scale might
have a cosmological density of astrophysical interest.

In the case of the supersymmetric relic $\chi$, detailed calculations 
have been performed \cite{relic}, and it has often been observed 
that $0.01 < \Omega_{\chi} < 1$ is a generic feature of
parameter choices in the minimal supersymmetric extension of the Standard
Model (MSSM). Moreover, it has also often been argued that restricting
$\Omega_{\chi} h^2 < 1$ suggests very strongly that $m_{\chi}$ is at
most a few hundred GeV \cite{uplim}. However, it is known that there are
rays in parameter space along which $\Omega_{\chi} h^2$ may be kept small
even though sparticle masses grow large. The purpose of this paper is to
bring together and complete these observations, with the aim of
clarifying the extent to which the supersymmetric resolution of the
fine-tuning problem is related to the cosmological fine-tuning
problem.

The criterion for hierarchical fine tuning that we use is that championed
in~\cite{FINETUNE,BAGI}, namely  the logarithmic sensitivity $\Delta_0$ of
$M_Z$ to variations in MSSM input parameters $a_i$:
\begin{equation}
\Delta_0 \equiv {\rm max} |\Delta_{a_i}|: \;\;\; \Delta_{a_i}={a_i\over
M_Z}{\partial M_Z\over\partial a_i}
\label{defineDelta0}
\end{equation}
The lower bounds imposed on $\Delta_0$ by experiments at LEP and
elsewhere have been discussed previously by several authors. There
have been extensive discussions how this price varies as a function
of $\tan\beta$, the ratio of Higgs vacuum expectation values (vev's),
and how the price may be reduced if some underlying theory
imposes relations between some MSSM parameters \cite{CEP,BAST,CEOP}.
Here we extend these discussions to include a cosmological dimension.

We study the extent to which the cosmological fine-tuning problem
is related to the fine tuning of the gauge hierarchy,
as measured by the quantity $\Delta_0$ (\ref{defineDelta0}).
Making an extensive search of the MSSM parameter space, we find that 
minimizing $\Delta_0$ leads to $\Omega_{\chi} h^2 \sim 0.1$. Moreover, 
we find a general correlation between those parameter choices with larger 
$\Omega_{\chi} h^2$ and those with larger $\Delta_0$. On the other hand,
imposing
$\Omega_{\chi} h^2 < 1$ does not exclude all models with large $\Delta_0$. 
We can distinguish three regions in the neutralino mass $m_\chi$ (or,
correspondingly, in the GUT scale gaugino mass parameter $M_{1/2}$) 
in which $\Omega_{\chi} h^2 < 1$ with large $\Delta_0$ are possible. For
$m_\chi\approx M_Z/2$  or $m_\chi\approx M_h/2$ (that is for
$M_{1/2}\simlt150$ GeV) large pole-dominated 
$s-$channel annihilation cross sections give a low relic density in ways 
unrelated to the value of $\Delta_0$. In the intermediate region, 
$M_Z/2$, $M_h/2\simlt m_\chi\simlt m_t$ neutralino annihilation proceeds
mainly through slepton exchange, and $\Omega_{\chi} h^2 < 1$ then implies 
$m_0/M_{1/2} \simlt {\cal O}(1)$ and small $\Delta_0$. For $m_\chi\simgt m_t$,
$\Omega_{\chi} h^2 < 1$ is also possible in models with relatively light
stop, so that the $t-$channel annihilation into $t\bar t$ pair is 
enhanced again leading to a lower relic density. Within the minimal SUGRA
framework such models require large left-right mixing in the stop sector 
and hence, in addition to a large top Yukawa coupling, a large value of
$A_0$. Thus, models with $\Omega_{\chi} h^2 < 1$ and large fine-tuning 
are rather special: some have $m_\chi\approx M_Z/2$ or $M_h/2$ whilst 
others have $m_\chi\simgt m_t$ and relatively light stop. 
The former class of special solutions can be exhaustively
explored by chargino searches at LEP~200, whilst the latter would be
absent if
$|A_0| \simlt 1$~TeV. Our study does not find that the cosmological and
hierarchical fine-tuning problems are equivalent, but it does confirm
that an interesting cosmological relic density is indeed to be expected in
supersymmetric models that do not exhibit extreme fine tuning.

Our study of these issues is based on the survey of MSSM parameter
space made in~\cite{CEP,CEOP}, which we review briefly here.
As usual, we denote the Higgs mixing parameter by $\mu$, and
we assume the conventional minimal parameterization of soft
supersymmetry breaking in the MSSM, via a universal scalar mass
parameter $m_0$, a universal gaugino mass parameter $M_{1/2}$, and a
trilinear (bilinear) coupling $A_0$ $(B_0)$. We assume that $\mu_0$, $m_0$,
$M_{1/2}$, $A_0$ and $B_0$ are the appropriate inputs $a_i$ 
at the GUT scale $M_{GUT} = 2 \times 10^{16}$~GeV which should be used in
the hierarchical fine-tuning criterion (\ref{defineDelta0}). 
These parameters are renormalized down to the
electroweak scale in the standard way, leading to an effective potential
that breaks electroweak gauge symmetry spontaneously, with a calculable
value of $\tan\beta$. 

Our procedure for surveying the MSSM parameter space is, for reasons of
convenience, to choose low-energy parameter sets 
that respect the experimental constraints and yield an appropriate
electroweak vacuum, as determined using the full one-loop
effective potential in the MSSM \cite{OLPO}. Specifically, using the
measured value of $M_Z$, for each value of
$\tan\beta$ and a given $sgn(\mu)$, we scan low-energy values of the
left-handed doublet squark mass
$m_Q$, the right-handed singlet up-squark mass $m_U$ and the CP-odd MSSM
Higgs mass $M_A$ that are allowed by the experimental
constraints~\cite{OLPO}. We then use the renormalization-group equations
to find the corresponding allowed values of the GUT input parameters.

As experimental constraints,
we take into
account the precision electroweak data published at the Moriond conference
\cite{LEPEWWG}, and  require $\Delta\chi^2<4$ in a global MSSM fit. We
also incorporate LEP lower limits on the masses of sparticles and Higgs
bosons \cite{LEPHiggs}.  Another important accelerator constraint is
provided by the recently  measured $b\rightarrow s\gamma$ branching ratio
$2\times10^{-4}< Br(B\rightarrow X_s\gamma)<4.5\times10^{-4}$
\cite{CLEO},  which we treat as described 
in~\cite{CEP}. We find that the GUT-scale parameters corresponding to
successful choices with an upper cut of 1.2 TeV on the soft masses  
$M_A$ and the squark masses, vary over the ranges $|\mu_0| \lappeq 2$~TeV,
$m_0 \lappeq 1.7$~TeV, $M_{1/2} \lappeq 600$~GeV, $|A_0| \lappeq 5$~TeV 
and $|B_0|\lappeq 3$~TeV. Note, in particular, that the upper bound on 
$M_{1/2}$ (and, in consequence, on the chargino mass) visible on the plots 
follows simply from the cut imposed on the scanning procedure.

Finally, we note that the range of the relic
neutralino density favoured by astrophysics and cosmology is
$\Omega_{\chi} h^2 \sim 0.1$ (see, e.g., \cite{efos2}), but we do not use
this as a constraint in our analysis. Rather, our aim will be to explore
the extent to which this is a natural outcome for successful
MSSM parameter choices with small values of the hierarchical
fine-tuning measure $\Delta_0$ (\ref{defineDelta0}).  To this end, within
the context of the supergravity-based MSSM, we calculate the relic
density for each of the models considered in the fine-tuning analysis of
\cite{CEP,CEOP}. 
Thus, for every allowed set of the GUT-scale parameters obtained by
our scanning procedure, we have a calculation of the relic density
$\Omega_{\chi} h^2$ in terms of all of the low-energy masses,
which are determined from $M_{1/2}, m_0$ and
$A_0$ for the same fixed values of $\tan \beta$ and $sgn(\mu)$. 

We note that there are two recent refinements of the analysis of
the MSSM parameter space and the dark matter density that 
have not been included
in this survey. One is the latest 
implementation~\cite{AF} of the requirement that our electroweak vacuum be
stable against possible transitions to vacua that violate
charge and/or color conservation. This requirement
tends to exclude parameter
choices with $m_0 / M_{1/2} \simlt 1/2$, which do not have exceptional
values of either $\Delta_0$ or $\Omega_{\chi} h^2$. We have also omitted
the possibility of coannihilation~\cite{EFO} between $\chi$ and the
${\tilde
\tau}_R$, which is the next-to-lightest supersymmetric particle in a
generic domain of parameter space. This is important when $m_{\chi}
\simlt 1.1 m_{{\tilde \tau}_R}$, which is the case only for a very small
number of the parameter choices in our survey. When it is significant,
it tends to enable points with larger hierarchical fine tuning $\Delta_0$
to have a cosmologically interesting value of $\Omega_{\chi} h^2$.

Results for the case tan$\beta = 2.5$ are shown in Fig.~1. The
top left panel displays directly the correlation we find between
the hierarchical fine-tuning parameter $\Delta_0$ and the neutralino
relic density. Here and elsewhere,  the eight-pointed stars
represent parameter choices where no specific direct-channel annihilation
mechanism is dominant. The five-pointed stars represent parameter choices
where
$m_{\chi}\sim M_Z/2$, so that $\chi \chi$ annihilation via
the direct-channel $Z^0$ pole is dominant. Because the width of the
$Z^0$ is relatively large, the effect of $s$-channel annihilation
through $Z^0$'s occurs only in a small patch of the parameter space. The
open circles represent parameter choices where
$m_{\chi} \sim M_h/2$, so that $\chi \chi$ annihilation via the
direct-channel pole of the lightest MSSM Higgs boson $h^0$ is dominant.
In this case, because of the very small width for $h^0$, the suppression
of the relic density actually begins even when $2m_\chi \sim 0.8 M_h$
\cite{gs} and thus can cover a broader parameter volume. Finally, the
dots represent parameter choices where
$m_{\chi} \gappeq m_t$, so that
$\chi \chi\rightarrow\bar t t$ annihilation is important.

We see in the top left panel of Fig.~1 that (i) the minimum value of
$\Delta_0 \sim 13$ is attained for $\Omega_{\chi} h^2 \sim 0.1$, (ii) 
the minimum value of $\Delta_0$ increases gradually for larger
values of $\Omega_{\chi} h^2$, (iii) apart from a few exceptional 
points, mainly with $m_\chi > m_t$, all choices with $\Omega_{\chi} h^2
<1$
have $\Delta_0 < 100$, and (iv) apart from choices where
$\chi \chi\rightarrow {\bar t} t$ dominates, when
$\Omega_{\chi} h^2 > 1$ there is a clear tendency for its
value to be correlated with that of $\Delta_0$. The 
origins of the exceptional
choices with low $\Omega_{\chi}h^2$ and large $\Delta_0$
are seen in the top right panel of Fig.~1. They have values
of $M_{1/2} \sim 100$ to $150$~GeV, corresponding to $m_{\chi} \sim
M_Z/2$ and/or $M_h/2$. We also see in this panel that the 
choices with annihilation into ${\bar t} t$ correspond
to $M_{1/2}\gappeq 400$~GeV, which are the largest in our sample.
It is therefore not surprising that these correspond to 
some of the largest values of $\Delta_0$, as we see in
the top left panel.

The remaining panels of Fig.~1 show how the value of
$\Omega_{\chi} h^2$ is correlated with the values of other
MSSM parameters. In the middle left panel, we see a clear
correlation between $\Omega_{\chi} h^2$ and the ratio $m_0/M_{1/2}$,
again apart from a few choices where $\chi \chi \rightarrow Z^0$ or
$h^0$
or $\chi \chi \rightarrow t\bar t$ 
dominates. Apart from these choices, we see that $\Omega_{\chi} h^2$
is minimized for $m_0 / M_{1/2} \lappeq 1$. 
In the middle right panel, we see that $\Omega_{\chi} h^2<1$ is consistent
with relatively small values of $|A_0|$ for the $m_\chi\approx M_Z/2$ or 
$M_h/2$ points and the points with no dominant annihilation channel, 
whereas large $A_0$ is required when $m_\chi>m_t$. This effect is even more
pronounced for larger values of $\tan\beta$ (see Fig.~2 and related
comments below).

It is worth emphasizing that the parameter choices dominated by
$\chi \chi \rightarrow Z^0$ annihilation will soon be explored
directly by LEP, as will to a large extent those parameters which lead to
annihilation via the light Higgs pole. These choices have $m_{\chi^{\pm}}
\lappeq 100$~GeV, and hence should reveal their secrets when the LEP
center-of-mass energy is increased to $\sim 200$~GeV in
the years 1999 and 2000.

We have carried out similar parameter studies for several
larger values of $\tan\beta\le30$, as well as for $\tan\beta = 1.65$,
and now discuss some similarities and differences in these cases.
As a general rule, the cases with $\tan\beta > 2.5$ are 
qualitatively similar
to the $\tan\beta = 2.5$ case. In particular, it is always true that
minimizing $\Delta_0$ favours $\Omega_{\chi} h^2 \sim 0.1$, as seen
in Fig.~2 for $\tan\beta = 10$, for example. 
On the other hand there is no trend for the
upper bound on $\Delta_0$ to be improved if one selects $\Omega_{\chi}
h^2 < 1$. We see again in the top left panel of Fig.~2 three distinctive
sets of parameter choices which satisfy this condition. The one
with $\chi \chi \rightarrow Z^0$ or $h^0$ annihilation,  is
unrelated to the values of $\Delta_0$, as seen in the top left panel. 
The case with no dominant annihilation channel has small $\Delta_0$. 
We also see there that the $\chi \chi \rightarrow\bar tt$ cases have
distinctively larger values of $\Delta_0$. 
For $\chi \chi \rightarrow\bar tt$ cases the correlation of 
$\Omega_{\chi}h^2 < 1$ with large negative $A_0$ or very large positive 
$A_0$ is quite pronounced and, as explained earlier, reflects the 
necessity of a light stop, i.e., of a large left-right mixing. The 
asymmetry in the $A_0$ values follows from the 
renormalization-group running of the
mixing parameter down to low energies, where it is
related to its GUT-scale counterpart
through the equation~\cite{COPW}:
\begin{eqnarray}
A_t=A_0(1-y) - {\cal O}(2)M_{1/2}
\end{eqnarray}
where $y$ is the ratio of the top Yukawa coupling to its infra-red 
quasi-fixed point value.
The correlation between $\Omega_\chi h^2$ and the ratio 
$m_0/M_{1/2}$ is less pronounced, and $m_0/M_{1/2}>1$ is possible for
$\Omega_{\chi}h^2 < 1$ even away from the exceptional cases mentioned above.

The situation is rather different for $\tan\beta = 1.65$, as seen in
Fig.~3. Here, we see in the top left panel that the tendency to 
favour low $\Omega_{\chi} h^2$ is less
marked, though $\Omega_{\chi} h^2 \sim 1$ is still preferred. 
We again see that parameter choices with $\Omega_{\chi} h^2 < 1$
normally have $\Delta_0 < 200$, with exceptions provided by choices
with important direct-channel $Z^0$ and $h^0$ poles, as seen in the top
right panel of Fig.~3. We recall that the five-pointed star choices
with $m_{\chi} \sim M_Z/2$ will be explored exhaustively by the LEP
runs at $E_{CM} = 200$~GeV. We also see in the panel f) of Fig.~3
that LEP Higgs searches will be able to verify or exclude this
possible value of $\tan\beta$: it predicts that $M_h \lappeq 96$~GeV,
whereas LEP~200 should have a reach extending beyond $M_h = 100$~GeV.

The results presented so far have been obtained with a scan over
$m_Q$, $m_U$ and $M_A$ up to 1.2 TeV. We see that the highest values 
of $M_{1/2}$ in this scan (and the corresponding values of $m_\chi$) are 
consistent with $\Omega_{\chi} h^2 < 1$, but at the expense of increasing
fine-tuning. It is interesting to extend the scan up to higher values 
of soft masses so that an absolute upper bound on $M_{1/2}$ ($m_\chi$)
is obtained from the requirement $\Omega_{\chi} h^2 < 1$. This bound is shown
in Fig.~4, where for $\tan\beta=10$ we show $\Omega_{\chi} h^2$ versus
$\Delta_0$ and $M_{1/2}$, scanning over $m_Q$, $m_U$ and $M_A$ values 
up to 8 TeV.
In Fig.~5a we plot $\Delta_0$ as a 
function of $m_{\chi}$ 
for models which give $\Omega_{\chi} h^2 < 1$. We observe that the 
bound for the heavier superpartner masses 
is weak, around 1~TeV, but is saturated only for large $\Delta_0$.

Let us now summarize the story so far. Minimizing the gauge hierarchy
fine-tuning parameter $\Delta_0$ favours values of $\Omega_{\chi} h^2$
close to the range favoured by astrophysicists and cosmologists.
Conversely, restricting $\Omega_{\chi} h^2 < 1$ favours models with
relatively low values of $\Delta_0$, with certain well-understood
exceptions, some of which may soon be probed by experiments at
LEP. 

The question
now arises how sensitive these observations are to the fine-tuning
criterion we have used. In a recent paper~\cite{CEOP}, we have studied
the consequences of postulating some linear relation between a pair
of the MSSM input parameters, and we now discuss their possible
implications for the cosmological fine-tuning problem.
Fig.~6 displays the implications of assuming a linear relation between
$\mu_0$ and $M_{1/2}$, for the specific case $\tan\beta = 2.5$. 
We see that the global minimum of
$\Delta_{M \mu}$ is significantly reduced, and that the preference
for
$\Omega_{\chi} h^2 \sim 0.1$ is maintained and even enhanced, as compared
with Fig.~1 where no parameter relation was assumed. We also see that the
patterns of correlations between values of $\Omega_{\chi} h^2$ and
$\Delta_0$ for three different 
sets of parameters are maintained, modulo an approximate overall
rescaling
in the values of $\Delta_0$.
The picture if a linear relation between $\mu_0$ and $A_0$ is assumed is
somewhat different, as displayed in Fig.~7. 
The minimum of $\Delta_{A\mu}$
is again reduced, as compared to the case with no parameter relation,
but there is no longer any preference for $\Omega_{\chi} h^2 < 1$:
indeed, the favoured value is $\Omega_{\chi} h^2 > 10$. 
Finally, returning to Figs.~5b and 5c we see, for $\tan\beta=10$, the
values of $\Delta_{M \mu}$ 
and $\Delta_{A\mu}$ for points satisfying $\Omega_{\chi} h^2<1$ 
as functions of $m_{\chi}$.
The values $\Delta_{M \mu}\simlt 10$ are now compatible with
$m_{\chi}\simlt500$ GeV.

We conclude by reiterating that there is a significant
correlation between the amount of hierarchical fine tuning and
the relic cold dark matter density in the MSSM. It is indeed
``natural'' that the supersymmetric relic particle have
$\Omega_{\chi} h^2 \sim 0.1$~to~$1$, which we consider to be
an attractive feature of this dark matter candidate, as
compared to massive neutrinos or the axion, whose densities
have no obvious reason to fall within this favoured range.
However, this cannot be regarded as a hard prediction of the MSSM.
Moreover, fine tuning is always a subjective argument, rather
than a hard-and-fast mathematical argument, and one would
immediately embrace any experimental discovery of even an
``unnatural'' dark matter particle. Nevertheless, we
find this correlation between cosmological and hierarchical fine tuning
an interesting supplementary argument in favour of supersymmetric
cold dark matter.

\vskip 0.3cm
\noindent {\bf Acknowledgments.}

\noindent We thank Marek Olechowski for collaboration at the early stage of
this project. We would also like to thank Toby Falk for many helpful
conversations. The work of P.H.Ch. and S.P. has been partly supported by
the Polish State Committee for Scientific Research grant 2 P03B 040 12
(1998-1999).  The work of K.A.O. was supported in part by DOE grant
DE--FG02--94ER--40823 at the University of Minnesota.

\newpage
\thispagestyle{empty}

\begin{figure}
\psfig{file=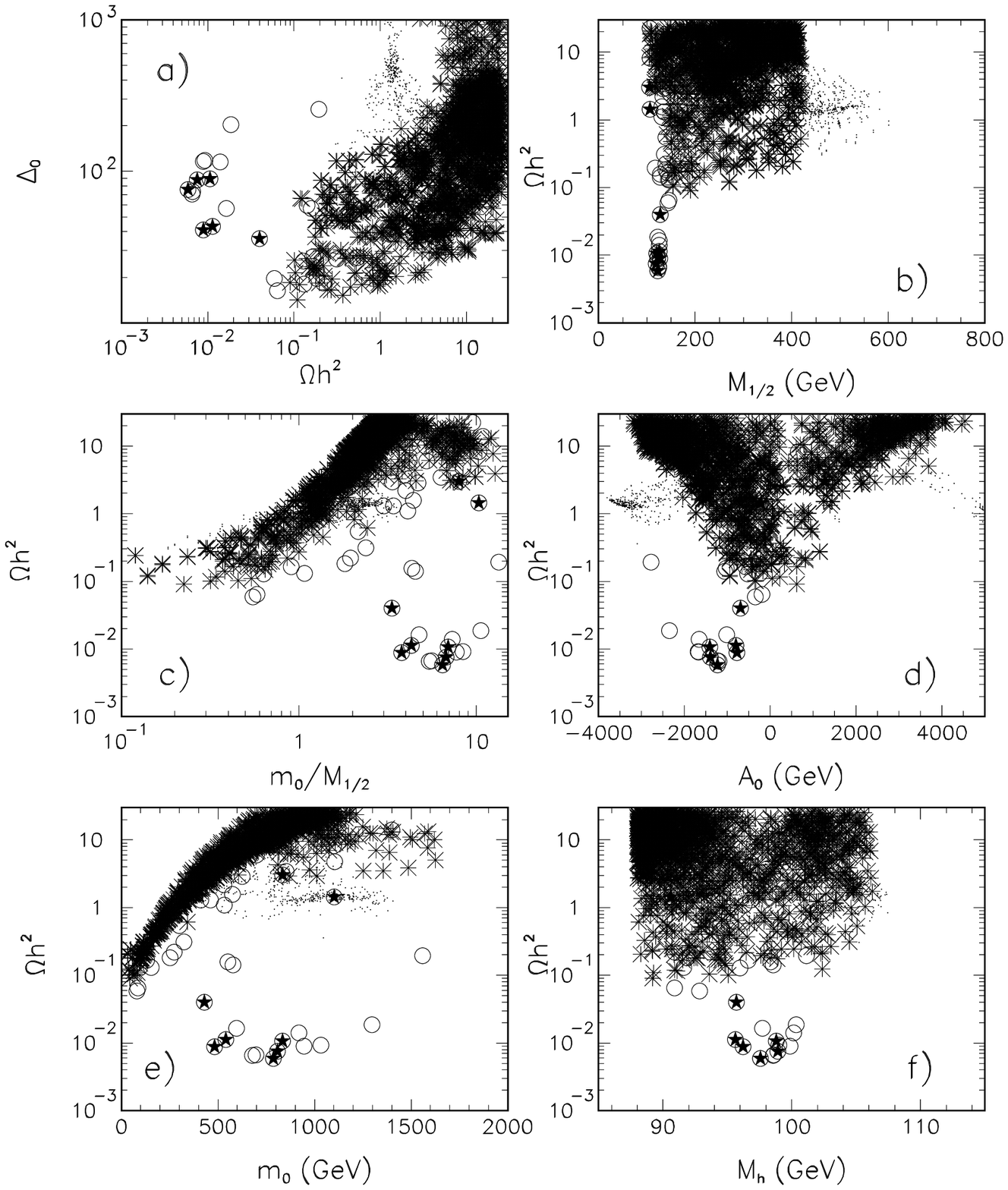,width=15.0cm,height=20.0cm}
\vspace{1.0truecm}
%height=13cm,bbllx=4.5cm,bblly=.cm,bburx=14.cm,bbury=13cm}}
\caption{{\it The price of fine tuning and $\Omega_\chi h^2$ for 
$\tan\beta=2.5$, as functions of various variables in the minimal 
supergravity model. The experimental constraints described in the text
are included. The five-pointed stars (open circles) represent cases where
$m_{\chi}\sim M_Z/2$ ($m_{\chi} \sim M_h/2$). The dots represent parameter
choices where $m_{\chi} \gappeq m_t$. Eight-pointed stars represent
parameter choices where no specific direct-channel annihilation mechanism
is dominant.}}
\end{figure}

\begin{figure}
\psfig{file=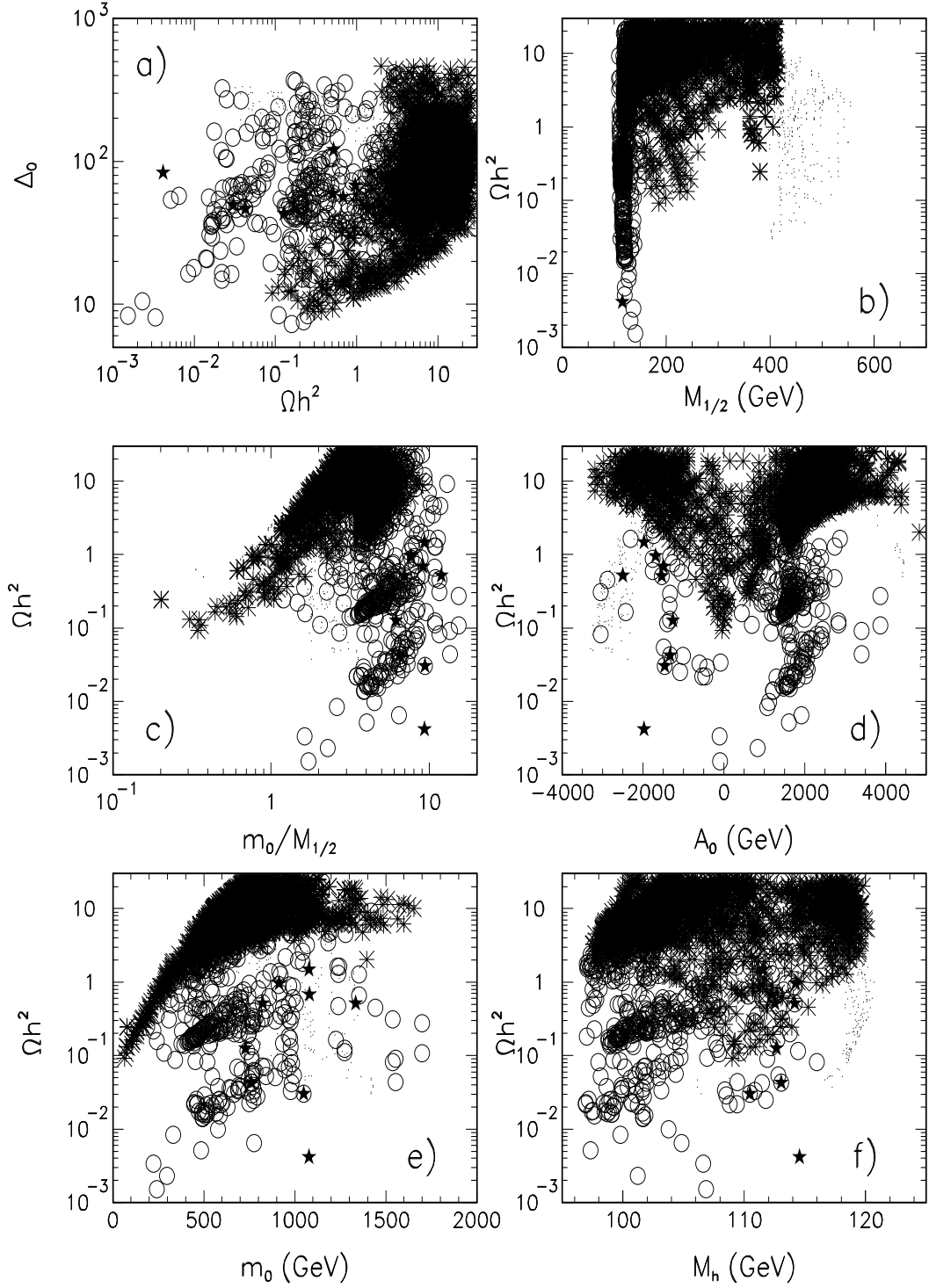,width=15.0cm,height=20.0cm}
\vspace{1.0truecm}
%height=13cm,bbllx=4.5cm,bblly=.cm,bburx=14.cm,bbury=13cm}}
\caption{{\it As in Fig.~1, but for $\tan\beta=10$.}}
\end{figure}

\begin{figure}
\psfig{file=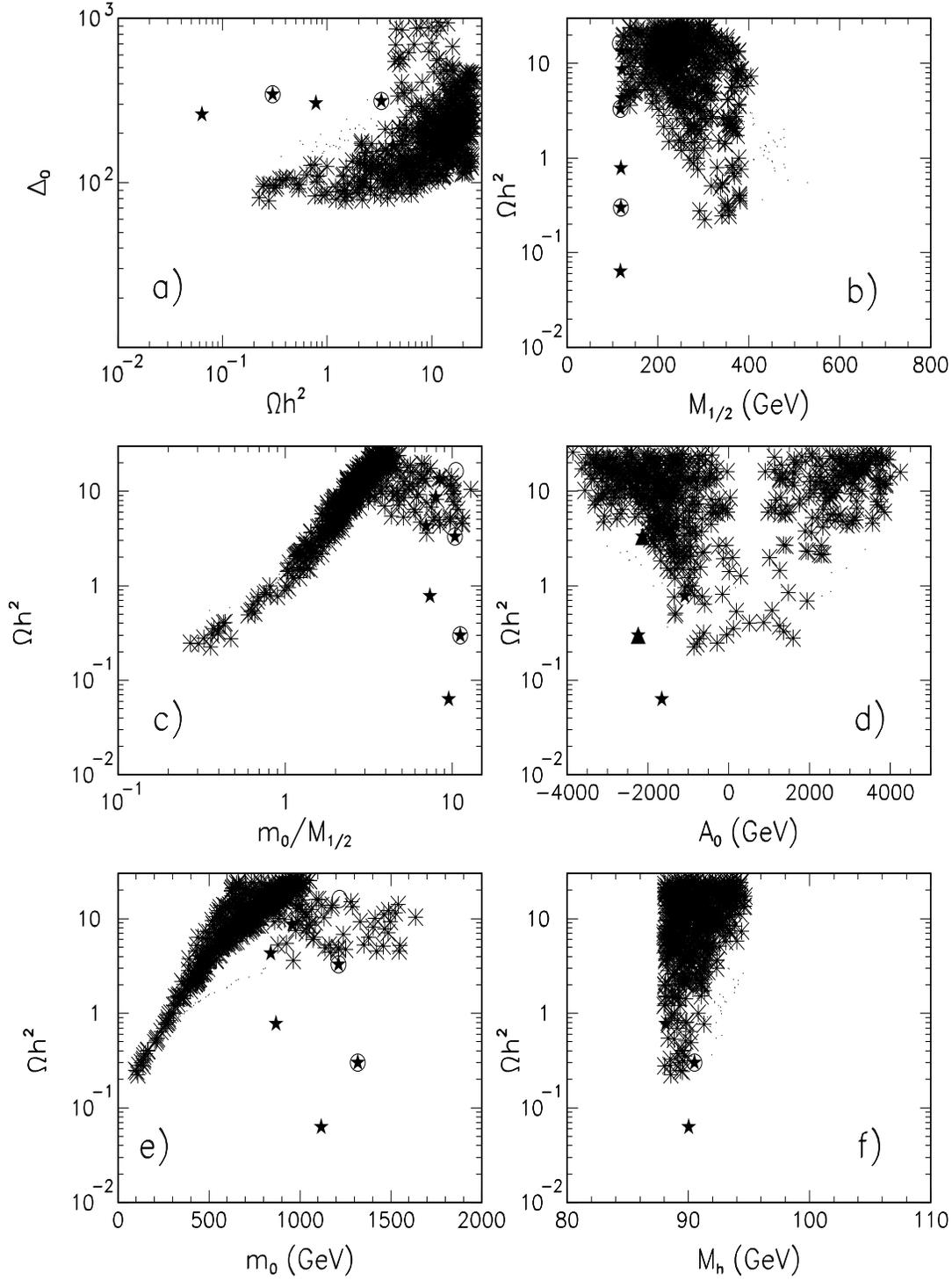,width=15.0cm,height=20.0cm}
\vspace{1.0truecm}
%height=13cm,bbllx=4.5cm,bblly=.cm,bburx=14.cm,bbury=13cm}}
\caption{{\it As in Fig.~1, but for $\tan\beta=1.65$.}}
\end{figure}

\newpage
\begin{figure}
\psfig{file=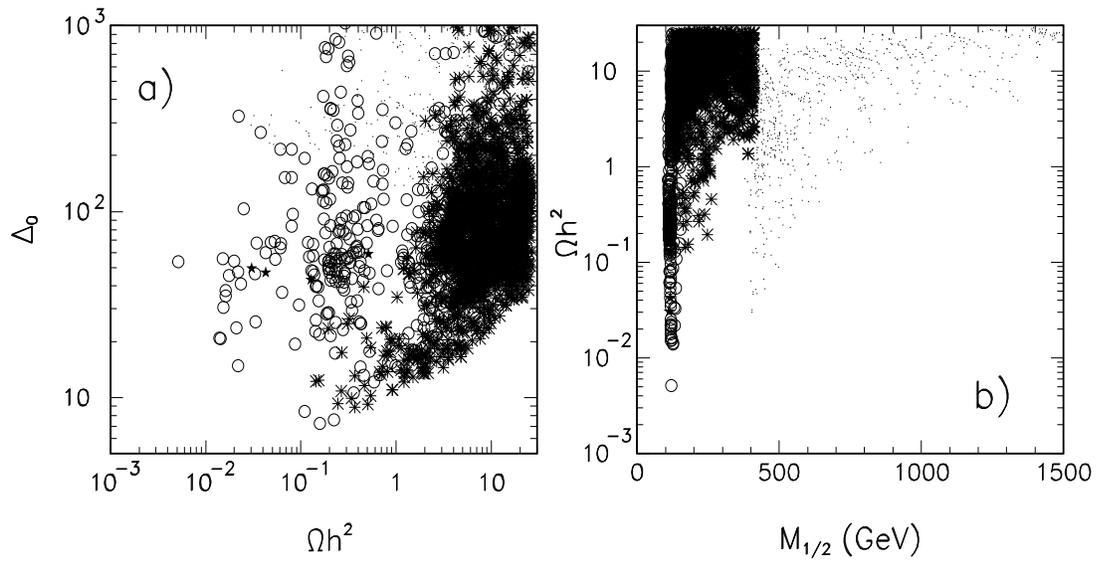,width=15.0cm,height=8.0cm}
\vspace{1.0truecm}
%height=13cm,bbllx=4.5cm,bblly=.cm,bburx=14.cm,bbury=13cm}}
\caption{{\it The price of fine tuning and $\Omega_\chi h^2$ for 
$\tan\beta=10$, with the scan over $m_Q$, $m_U$ and $M_A$ extended 
up to 8 TeV.}}
\end{figure}

\newpage
\begin{figure}
\psfig{file=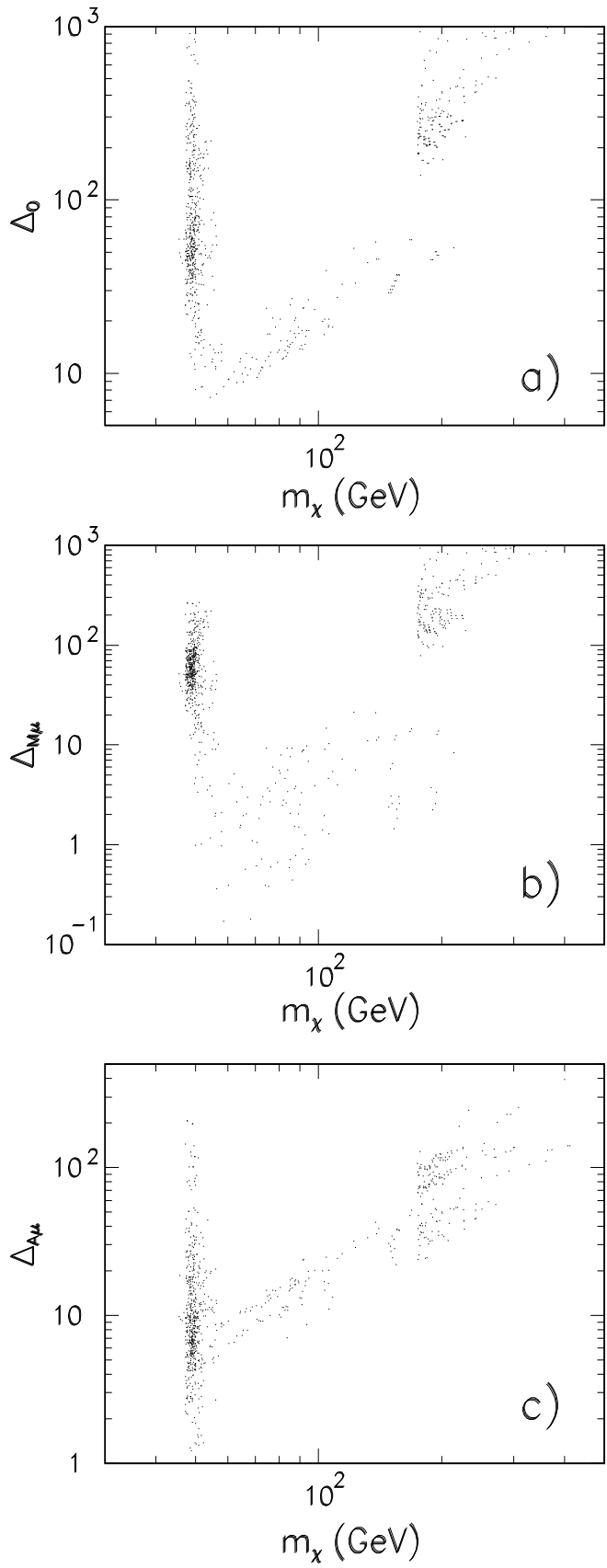,width=15.0cm,height=20.0cm}
\vspace{1.0truecm}
%height=13cm,bbllx=4.5cm,bblly=.cm,bburx=14.cm,bbury=13cm}}
\caption{{\it The fine-tuning measures
$\Delta_0$, $\Delta_{M\mu}$ and $\Delta_{A\mu}$ as functions
of the lightest neutralino mass for $\tan\beta=10$. Only points
satisfying $\Omega_\chi h^2<1$ are shown.}}
\end{figure}

\begin{figure}
\psfig{file=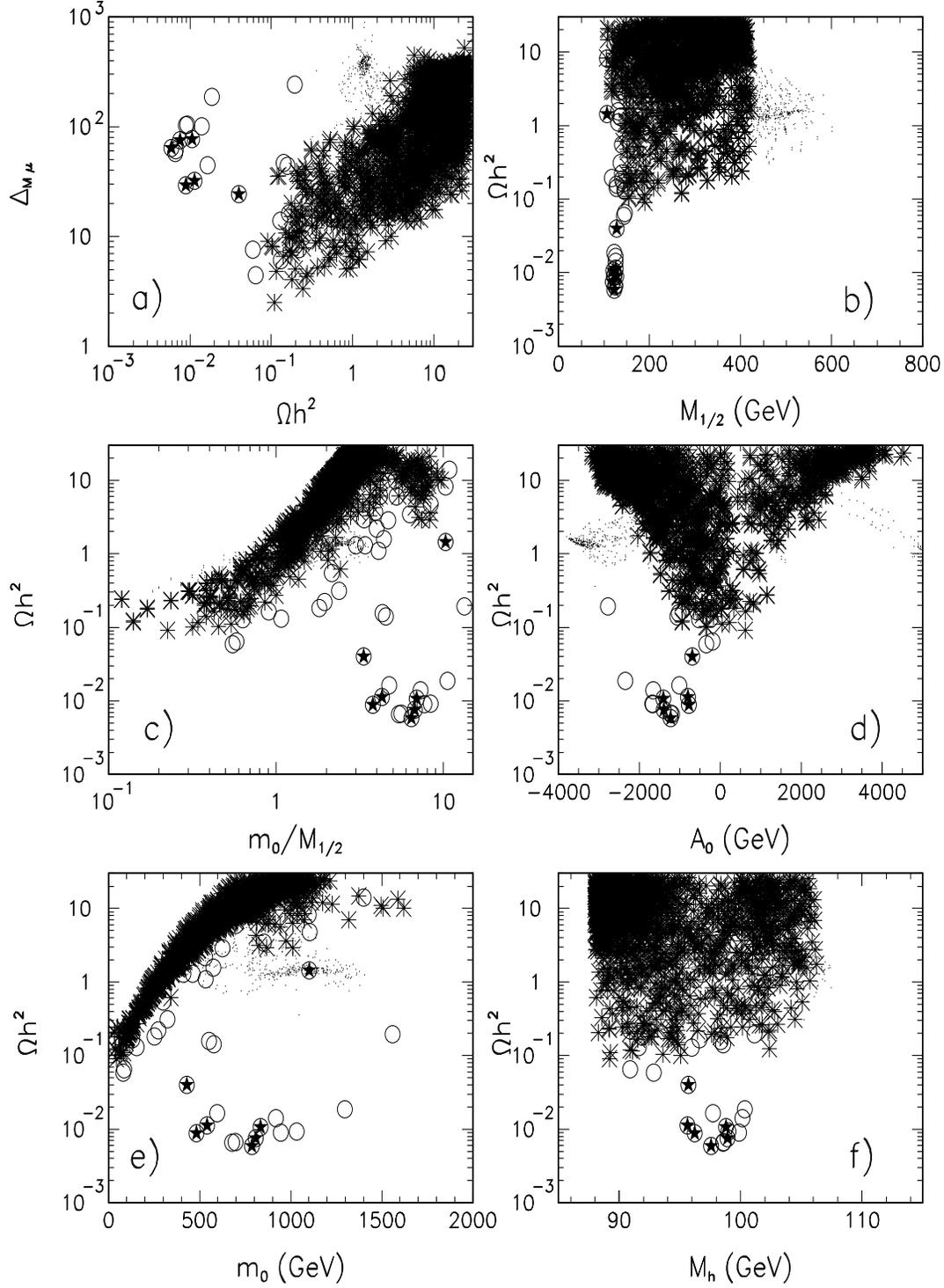,width=15.0cm,height=20.0cm}
\vspace{1.0truecm}
%height=13cm,bbllx=4.5cm,bblly=.cm,bburx=14.cm,bbury=13cm}}
\caption{{\it As in Fig.~1 ($\tan\beta=2.5$), but assuming a
linear correlation between $M_{1/2}$ and $\mu_0$. We show only points
with $\Delta_{M \mu}<\Delta_0$.}}
\end{figure}

\begin{figure}
\psfig{file=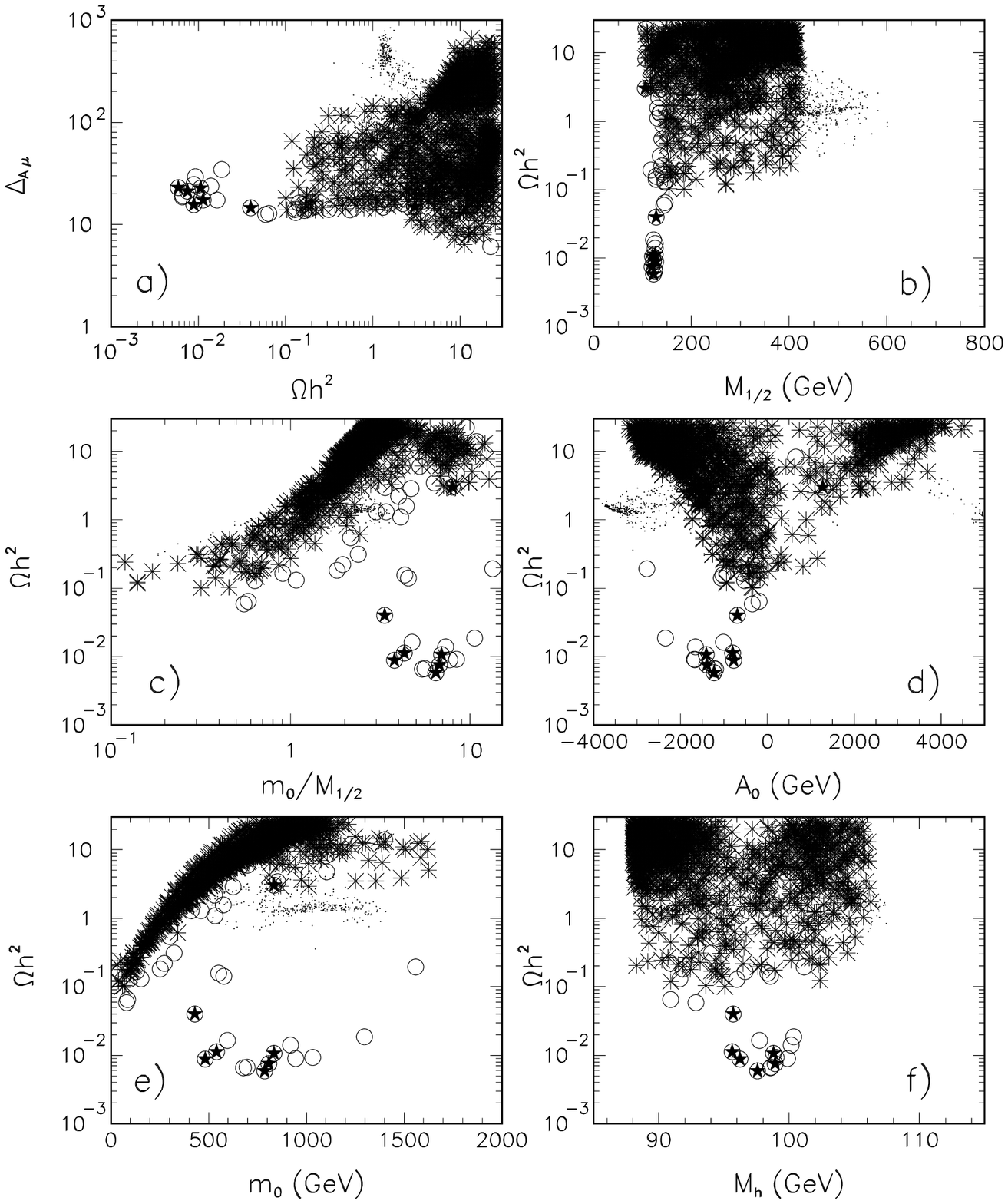,width=15.0cm,height=20.0cm}
\vspace{1.0truecm}
%height=13cm,bbllx=4.5cm,bblly=.cm,bburx=14.cm,bbury=13cm}}
\caption{{\it As in Fig.~1 ($\tan\beta=2.5$), but instead assuming a
linear correlation between $A_0$ and $\mu_0$. We show only
points with $\Delta_{A\mu}<\Delta_0$.}}
\end{figure}


\begin{thebibliography}{99}

\bibitem{mark} J. Ellis, J.S. Hagelin, D.V. Nanopoulos, K.A. Olive
and M. Srednicki, {\sl Nucl. Phys.} {\bf B238} (1984) 453; \\
M. Srednicki, for the Particle Data Group, 
{\sl Eur. Phys. J.} {\bf C3} (1998) 125.

\bibitem{raffelt} G. Raffelt, {\sl Phys. Rep.} {\bf 198} (1990) 1 and
                  astro-ph/9707268; \\ 
                  M.S. Turner, {\sl Phys. Rep.} {\bf 197} (1990) 67.

\bibitem{jkg} G. Jungman, M. Kamionkowski and K. Griest, {\sl Phys.
              Rep.} {\bf 267} (1996) 195.

\bibitem{naturalness}L. Maiani, {\sl Proceedings Summer School on Particle
        Physics}, Gif-sur-Yvette, 1979 (IN2P3, Paris, 1980), p.3;\\
        G. 't Hooft, in {\sl Recent Developments in Field
        Theories}, eds. G. 't Hooft et al., (Plenum Press, New York,
        1980);\\  
        E. Witten, {\sl Nucl. Phys.} {\bf B188} (1981) 513;\\
        R.K. Kaul, {\sl Phys. Lett.} {\bf 109B} (1982) 19.

\bibitem{SGUT} J. Ellis,  S. Kelley and D.V. Nanopoulos, {\sl Phys. Lett.} 
               {\bf B249} (1990) 441 and {\bf B260} (1991) 131;\\ 
               C. Giunti, C.W. Kim and U.W. Lee, {\sl Mod. Phys. Lett.}
               {\bf A6} (1991) 1745;\\
               U. Amaldi, W. de Boer and H. F\"urstenau, {\sl Phys. Lett.}
               {\bf B260} (1991) 447;\\
               P. Langacker and M. Luo, {\sl Phys. Rev.} {\bf D44} (1991) 817.

\bibitem{CHPLPO} P.H. Chankowski, Z. P\l uciennik and S. Pokorski,
                 {\sl Nucl. Phys.} {\bf B439} (1995) 23.

\bibitem{hundred} M. Gr\"unewald and D. Karlen, talks at XXIX {\it
		International Conference on High-Energy
               Physics} Vancouver, B.C., Canada,
		{ \tt http://www.cern.ch/LEPEWWG/misc/}.

\bibitem{susyhiggsmass} M. Carena, M. Quiros and C.E.M. Wagner,
{\sl Nucl. Phys.} {\bf B461} (1996) 407;\\
H.E. Haber, R. Hempfling and A.H. Hoang, {\sl Zeit.
f\"ur Phys.} {\bf C75} (1997) 539.

\bibitem{Dim} S. Dimopoulos, {\sl Phys. Lett.} {\bf B246}
(1990) 347.

\bibitem{relic} J. McDonald, K.A. Olive and M. Srednicki, {\sl Phys.
                Lett.} {\bf B283} (1992) 80;\\
                M. Drees and M.M. Nojiri, {\sl Phys. Rev.} {\bf D47}
                (1993) 376.

\bibitem{uplim} K.A. Olive and M. Srednicki, {\sl Phys. Lett.} {\bf B230}
                (1989) 78 and {\sl Nucl. Phys.} {\bf  B355} (1991) 208; \\
                K. Griest, M. Kamionkowski and M.S. Turner, Phys. Rev.
                {\bf D41} (1990) 356;\\
                R. Arnowitt and P. Nath, {\sl Phys. Rev. Lett.} {\bf 70} 
                (1993) 3696; \\
                G. Kane, C. Kolda, L. Roszkowski and J. Wells, 
                {\sl Phys. Rev.} {\bf D49} (1994) 6173; \\
                T. Falk and K.A. Olive,  Phys. Lett. {\bf B375} (1996) 196.

\bibitem{FINETUNE} J. Ellis, K. Enqvist, D.V. Nanopoulos and F. Zwirner,
                   {\sl Nucl. Phys.} {\bf B276} (1986) 14.

\bibitem{BAGI} R. Barbieri and G.-F. Giudice, {\sl Nucl. Phys.} {\bf B306} 
               (1988) 63;\\
               S. Dimopoulos and G.-F. Giudice, {\sl Phys. Lett.} {\bf
               B357} (1995) 573.

\bibitem{CEP} P.H. Chankowski, J. Ellis and S. Pokorski {\sl Phys. Lett.}
              {\bf B423} (1998) 327.

\bibitem{BAST} R. Barbieri and A. Strumia, {\sl Phys. Lett.} {\bf B433}
               (1998) 63.

\bibitem{CEOP} P.H. Chankowski, J. Ellis, M. Olechowski and S. Pokorski,
               preprint CERN-TH-98-119 (hep-ph/9808275).

\bibitem{OLPO} M. Olechowski and S. Pokorski, {\sl Nucl. Phys.} {\bf B404}
               (1993) 590.

\bibitem{LEPEWWG} D. Reid, talk at XXXIII {\it Rencontres de Moriond}
                  (Electroweak Interactions and Unified Theories), Les Arcs, 
                  France, March 1998;
                  LEP Electroweak Working Group, report LEPEWWG/98-01.

\bibitem{LEPHiggs} ALEPH Collaboration, ALEPH 98-029 CONF 98-017;\\
                   DELPHI Collaboration, talk by K. Moenig, LEPC Meeting, 
                   CERN, March 31, 1998;\\
                   M. Acciarri (L3 Collaboration) et al., preprint 
                   CERN-EP/98-052; \\
                   OPAL Physics Note PN340, March 1998;\\
                   V. Ruhlmann-Kleider, talk at the XXXIII
                   {\it Rencontres 
                   de Moriond} (QCD and high energy interactions), Les Arcs, 
                   France, March 1998; \\
                   M. Felcini, talk at the 1st European Meeting {\it From
                   Planck Scale to Electroweak Scale}, Kazimierz, Poland, 
                   May 1998.

\bibitem{CLEO} CLEO Collaboration, preprint CLEO CONF 98-17 submitted to 
               the XXIX {\it International Conference on High-Energy 
               Physics} Vancouver, B.C., Canada, contributed paper ICHEP98
1011. 

\bibitem{efos2} J. Ellis, T. Falk, K.A. Olive and M. Schmitt,
                {\sl Phys. Lett.} {\bf B413} (1997)  355.

\bibitem{AF} S. Abel and T. Falk, CERN preprint TH/98-322
(hep-ph/9810297).

\bibitem{EFO} J. Ellis, T. Falk and K.A. Olive, CERN preprint TH/98-326
(hep-ph/9810360).

\bibitem{gs} K. Griest and D. Seckel, {\sl Phys. Rev.} {\bf D43} (1991) 3191. 

\bibitem{COPW} M. Carena, M. Olechowski, S. Pokorski and C.E.M. Wagner, 
{\sl Nucl. Phys.} {\bf B419} (1994) 213.

\end{thebibliography}
\end{document}